\documentclass[a4paper,11pt]{article}
\usepackage{authblk}
\usepackage{graphicx} 
\usepackage{cite} 
\usepackage{hyperref}
\usepackage{amsmath}
\usepackage{amssymb} 
\usepackage{subfig}
\usepackage{multirow}
\usepackage{fancyhdr}
\usepackage{amsmath}
\usepackage{adjustbox}
\usepackage[headheight=14pt]{geometry}
\usepackage{epstopdf} 
\usepackage{rotating} 

\title{Thermal Effects on the Moment of Inertia and Gravitational Redshift of PSR J1012+5307: Implications for Hyperonic Matter under SU(3) and SU(6) Symmetries\thanks{Corresponding authors. E-mail: xuy@cho.ac.cn, shenyf@cho.ac.cn, ziyu$_{-}$njfu@163.com} 
\thanks{This work was supported by the Development Project of Science and Technology of Jilin Province (Grant No. 20250102012JC), the Astrometric Reference Frame project (Grant No. JZZX-04), the Special Project for the Theoretical Basic Research of Changchun Satellite Observatory, National Astronomical Observatories, Chinese Academy of Sciences (Grant No. Y990000205), the National Natural Science Foundation of China (Grant No. 12465023), Guangxi Natural Science Foundation (Grant No.2025GXNSFAA069452), and Project for Enhancing the Basic Scientific Research Capability of Young and Middle-aged Teachers in Guangxi Colleges and Universities (Grant No. 2025KY0925).}}

\author[a,d]{Y. Xu}
\author[a]{X. L. Huang}
\author[a] {Y. B. Wang}
\author[a]{Q. Yuan}
\author[b]{W. B. Ding}
\author[a]{N. An}
\author[a]{Y. F. Shen}
\author[c]{Z.Yu}

\affil[a]{Changchun Observatory, National Astronomical Observatories, Chinese Academy of Sciences, Changchun 130117, China}
\affil[b]{College of Physics and Electronic Information Engineering, Guangxi Minzu Normal University, Chongzuo 532200, China}
\affil[c]{College of Science, Nanjing Forestry University, Nanjing 210037, China}
\affil[d]{School of Astronomy and Space Sciences, University of Chinese Academy of Sciences, Beijing 100049, China}

\date{12.05.2026}

\fancypagestyle{plain}{ 
    \fancyhf{}
    \fancyhead[C]{Submitted to Chinese Physics C}
}
\begin{document}

\maketitle
The temperature dependence of neutron star structure significantly alters the equation of state, thereby affecting observable properties such as the moment of inertia and gravitational redshift. Utilizing the relativistic mean-field (RMF) theory with hyperonic degrees of freedom under SU(3) flavor and SU(6) spin-flavor symmetries, we investigate the thermal effects on the structural properties of protoneutron stars (PNSs) and cold neutron stars (CNSs). Focusing on PSR J1012+5307, we analyze the drastic structural transformations occurring during the transition from a PNS to a CNS. For a 1.94 $M_{\odot}$ hyperonic star under SU(3) flavor symmetry, decreasing the temperature from $T = 30$ MeV  to 0 MeV induces a radius contraction of approximately 50$\% $, accompanied by a drop in the moment of inertia by nearly 26$\%$ and a significant increase​ in gravitational redshift by approximately 154$\%$. Furthermore, we examine the variations in the moment of inertia and gravitational redshift arising from mass uncertainties of PSR J1012+5307. Taking SU(3) flavor symmetry at $T=20$ MeV as an example, increasing the mass across the range 1.72 $M_{\odot}$$-$$1.94 M_{\odot}$ results in a radius contraction of 2.749 km, an $\sim$8\% increase in the moment of inertia, and a significant $\sim$40\% increase in the gravitational redshift. We find that in the cold regime and at a fixed mass, the radius, moment of inertia, and gravitational redshift of hyperonic matter under SU(3) flavor symmetry differ only marginally from those of purely nucleonic matter, rendering it difficult to observationally confirm the presence of hyperons in the core of PSR~J1012+5307.​ Moreover, future observations capable of precisely constraining pulsar masses—ideally through long-term monitoring from birth—hold the potential to determine more conclusively whether hyperons or other exotic matter reside in individual pulsars.


\section{Introduction}
Protoneutron Star (PNS) is a short-lived compact object formed by the core collapse and subsequent supernova explosion of massive stars at the late evolutionary stage. As a crucial evolutionary phase connecting supernova outbursts and cold neutron stars (CNSs), the internal structure and dynamical evolution of PNS directly determine the propagation and revival mechanism of supernova shocks, and regulate the heavy-element nucleosynthesis process as well as the time-varying characteristics of neutrino spectra\cite{Mezzacappa1998}. Meanwhile, the rotational evolution, magnetic field topology and gravitational deformation of PNS act as important physical origins for driving supernova gravitational wave radiation, constructing the central engine of short gamma-ray bursts, and dominating the evolutionary process of binary compact star mergers\cite{Ferrari2004,Rea2010,Lander2021,Margalit2022,Camelio2017}. Consequently, PNSs play an indispensable role throughout the whole life cycle of compact stars.

The interior of a PNS is characterized by extreme physical conditions, including ultra-high density, ultra-high temperature, strong gravitational fields, and efficient neutrino trapping. These environments far exceed the simulation capabilities of terrestrial laboratories, making the PNS a natural cosmic laboratory for investigating fundamental physical problems such as the equation of state (EOS) of hot and dense nuclear matter, the dynamics of strong interactions, superfluid/superconducting behaviors, and quark matter phase transitions \cite{Pons1997,Shen1998,Pons1999,Arcones2008}. The EOS describes the constitutive relationship among pressure, density, temperature, and composition of dense matter, serving as the key input that determines the structure and evolutionary behavior of PNSs\cite{Glendenning1991,Lattimer1991,Prakash1997,Glendenning2000,Bednarek2006,Yu20091,Yu20092,Roberts2012,Hong2016,Hong2018,Zhao20221,Zhao20222,HuoJL2026}. Different nuclear matter models, symmetry energy parameters, hyperon coupling strengths, and the inclusion or exclusion of quark degrees of freedom can lead to significant variations in the EOS. These differences further affect the mass–radius relation, the maximum stable mass, the critical rotation frequency, and the neutrino radiation characteristics of PNSs \cite{Glendenning1991,Prakash1997,Yu20091,Yu20092,Roberts2012}. Therefore, systematic investigations of the PNS EOS not only provide strong constraints on the properties of nuclear matter under extreme conditions and offer critical benchmarks for testing and refining existing nuclear physics models, but also facilitate a precise characterization of the macroscopic physical properties of PNSs. This, in turn, provides reliable theoretical support for deepening our understanding of supernova explosion mechanisms and for advancing frontier observations in gravitational-wave detection, neutrino astronomy, and time-domain surveys\cite{Duan2011,Camelio2016, Pascal2022,Bakir2026,Muller2026}.

However, there is currently a lack of systematic studies on how PNS matter containing a significant amount of hyperons can support the evolutionary process of intermediate-mass or massive pulsars—such as PSR J1012+5307, which has a measured mass of 1.83$\pm
$0.11$ M_{\odot}$\cite{Antoniadis2016,Mata2021,Wei2024}. In this work, we employ the relativistic mean field theory (RMFT) to investigate the moment of inertia and gravitational redshift of the PNS corresponding to the typical-mass PSR J1012+5307. Calculations are performed for matter including hyperons (npH matter), adopting the GM1 parameter set and incorporating SU(3) flavor and SU(6) spin-flavor symmetries. We aim to analyze the effects of temperature on the moment of inertia and gravitational redshift of PSR J1012+5307, with particular emphasis on the observational signatures arising from the PNS-to-CNS transition. This study seeks to place constraints on the EOS under extreme conditions and to explore the feasibility of probing exotic matter via future astrophysical observations.

The remainder of the work is organized as follows. In Section 2, we describe the finite-temperature, hyperon-included EOS for PNSs constructed within the RMFT framework, and present the formalism for calculating the stellar mass, moment of inertia, and gravitational redshift. Section 3 presents a detailed investigation of the moment of inertia and gravitational redshift of PNSs. Employing the GM1 parameter set within the RMFT framework, we study the effects of temperature on these properties for PSR J1012+5307 under both SU(3) flavor and SU(6) spin-flavor symmetries. Section 4 summarizes the main results.
\section{Theoretical Framework}
\subsection{The EOS of PNS in the RMFT}
Within the RMFT framework, PNS matter is described by the following Lagrangian density\cite{Yu20091,Hong2016,Zhao20221,Zhao20222,HuoJL2026}:
\begin{equation}
\begin{aligned}
\mathcal{L} 
&= \sum_B \bar{\Psi}_B\bigl(i\gamma_\mu\partial^\mu - m_B 
   - g_{\omega B} \gamma_\mu \omega^\mu 
   - g_{\phi B} \gamma_\mu \phi^\mu 
   - g_{\rho B} \gamma_{\mu} \vec{\rho}^\mu \cdot \vec{I}_{B} 
   + g_{\sigma B}\sigma + g_{\sigma^* B}\sigma^*\bigr)\Psi_B \\
&\quad + \sum_{l=e} \bar{\Psi}_l(i\gamma_\mu\partial^\mu - m_l)\Psi_l 
   +\frac{1}{2} m_\omega^2 \omega_\mu \omega^\mu 
   +\frac{1}{2} m^2_{\phi} \phi_\mu \phi^\mu 
   +\frac{1}{2} m_\rho^2 \vec{\rho}_\mu \cdot \vec{\rho}^\mu 
   - \frac{1}{3}a\sigma^3 
   - \frac{1}{4}b\sigma^4\\
&\quad  +\frac{1}{2}(\partial_v\sigma^*\partial^v\sigma^*-m^2_{\sigma^*}\sigma^{*2})
   + \frac{1}{2} (\partial_\mu \sigma \partial^\mu \sigma - m_\sigma^2 \sigma^2) 
   +\frac{1}{4}c_3 (\omega_\mu \omega^\mu)^2\\
&\quad - \frac{1}{4} W^{\mu v} W_{\mu v} 
   - \frac{1}{4} P^{\mu v} P_{\mu v}
   - \frac{1}{4} \vec{R}^{\mu v} \cdot \vec{R}_{\mu v} .
\end{aligned}
\end{equation}
In this context,  the baryon octet (p, n, $\Lambda$, $\Sigma^{+}$, $\Sigma^{0}$, $\Sigma^{-}$, $\Xi^{-}$, $\Xi^{0}$) and the leptons (e, $\nu_{e}$) are included, and PNS matter composed of them is referred to as npH matter. $\psi_{B(l)}$ ​denotes the Dirac field for baryons (or leptons), and $\gamma_{u}$ represents the Dirac matrices. The symbol $\vec{I}_{B}$ stands for the baryonic isospin matrix, while $m_B(l)$ corresponds to the mass of the respective baryon (or lepton). The coupling strengths between baryons and the vector mesons are characterized by the constants $g_{\omega B}$,  $g_{\phi B}$ and $g_{\rho B}$ for the $\omega$,  $\phi$ and $\rho$  mesons, respectively.  Additionally, the field-strength tensors for the three mesons are defined as  $W_{\mu v}=\partial_\mu\omega_v-\partial_v\omega_\mu$,  $P_{\mu v}=\partial_\mu\phi_v-\partial_v\phi_\mu$,$R_{\mu v}=\partial_\mu{\mathbf{\rho}}_v-\partial_v{\mathbf{\rho}}_\mu$. 

According to statistical physics, the partition function of the grand canonical ensemble is defined as\cite{Hong2016, Hong2018, HuoJL2026}
\begin{equation}
\begin{aligned}
Z = Tr\left[e^{-(\hat{H}-\mu_{B,l}\hat{N})/(K_{B}T)}\right].
\end{aligned}
\end{equation}
Here, $\hat{H}$ and $\hat{N}$ represent the Hamiltonian and particle number operators, respectively, with $K_{B}$ denoting the Boltzmann constant. In natural units, $K_{B}=1$. $T$ is the temperature. $\mu_{B,l}$ denote the chemical potential of baryons and leptons. They are related by
\begin{equation}
\begin{aligned}
\mu_{B,l}=\mu_n - q_i (\mu_e - \mu_{\nu e}),
\end{aligned}
\end{equation}

And the grand thermodynamic potential for nuclear matter of volume $V$ can be derived from the partition function
\begin{equation}
\begin{aligned}
\Omega_{gtp} = PV = -T \ln Z
\end{aligned}
\end{equation}
Within the RMF framework, for the Lagrangian given by Eq. (1), the Hamiltonian operator is
\begin{equation}
\begin{aligned}
\hat{H} = \int_V d^3x \left[ -\langle L \rangle + \bar{\Psi}\gamma_0 k_0 \Psi \rangle \right]
= -\langle L \rangle V + \sum_B \left[ \varepsilon_B \hat{N}_B + \overline{\varepsilon_B} \hat{\overline{N}}_B \right].
\end{aligned}
\end{equation}
Substituting Eq. (4) into Eq. (2), we can obtain the particle density $\rho_{B, l}= \frac{T}{V} \frac{\partial \ln Z}{\partial \mu_{B,l}}$, energy density $\varepsilon = \frac{T^2}{V} \frac{\partial \ln Z}{\partial T} + \mu_{B,l}\rho_{B, l}$,  pressure $P = \frac{T}{V} \ln Z$,  and entropy $S = \frac{1}{V} \left( \frac{\partial \Omega_{gtp}}{\partial T} \right)_{V, \mu}$, respectively.

Taking the natural logarithm of the partition function,  the partition functions for baryons and leptons can be expressed as follows:
\begin{equation}
\begin{aligned}
\ln Z_{B,l} = \frac{V}{T}\langle\mathcal{L}\rangle + \sum_{B,l} \frac{2J_{B,l}+1}{2\pi^2} \int_0^\infty k_{B,l}^2 dk_{B,l} \left\{\ln\left[1 + e^{-(\varepsilon_{B,l}(k_{B,l})-\mu_{B,l})/T}\right]\right\}.
\end{aligned}
\end{equation}
Here, $\varepsilon_{B,l}(k_{B,l})=\sqrt{k_{B,l}^2 + m_{B,l}^{*2}}$ is the thermal excitation energy of baryons and leptons.

Within the framework of the RMFT, the effective two-body nucleon–nucleon interaction is described by the exchange of meson fields. In this approach, the meson fields are replaced by their expectation values, and the meson masses and coupling constants are treated as free parameters, which are determined by fitting experimental data or established theoretical models. Consequently, the five mean meson fields satisfy the equations of motion
\begin{equation}
\begin{aligned}
\sum_B g_{\sigma B}\rho_{SB}=m_\sigma^2\sigma^0+a(\sigma^{0})^2+b(\sigma^0)^3,\\
\sum_B g_{\sigma^* B}\rho_{SB}=m_{\sigma^{*0}}^2\sigma^{*0},\\
\sum_B g_{\omega B}\rho_{B}=m_\omega^2\omega^0+c_{3}(\omega^{0})^{3},\\
\sum_B g_{\rho B}\rho_{B}I_{3B}=m_{\rho}^2\rho^{0}_{3},\\
\sum_B g_{\phi B}\rho_{B}=m_\phi^2\phi^0.
\end{aligned}
\end{equation}
Here, $\rho_{SB}$ and $\rho_{B}$ represent the scalar and vector densities associated with baryon B
\begin{equation}
\begin{aligned}
\rho_{SB} = \frac{1}{\pi^2} \int_{0}^{\infty} dk k^2 \frac{m_B^*}{\sqrt{k^2 + m_B^{*2}}}(\frac{1}{1 + exp[(\varepsilon_{B}(k) - \mu_{B})/T]}),\\
\rho_{B} = \frac{1}{\pi^2} \int_{0}^{\infty} dk k^2 (\frac{1}{1 + exp[(\varepsilon_{B}(k) - \mu_{B})/T]}).
\end{aligned}
\end{equation}
Based on this framework, the EOS for PNS matter can be expressed as follows, and the detailed derivation can be found in the references\cite{Glendenning2000,Bednarek2006, Yu20091,Hong2016,Hong2018,Zhao20221,Zhao20222}.
\begin{equation}
\begin{aligned}
\varepsilon &= \frac{1}{2} m_{\sigma}^{2} \sigma^{2} + \frac{1}{2} m_{\sigma^{*}}^{2} \sigma^{*2} + \frac{1}{2} m_{\omega}^{2} \omega_{0}^{2}+ \frac{1}{2} m_{\phi}^{2} \phi^{2} + \frac{1}{2} m_{\rho}^{2} \rho_{03}^{2}+ \frac{1}{3} a \sigma^{3} + \frac{1}{4} b \sigma^{4}+\frac{3}{4}c_{3}\omega^{4} \\
&\quad + \sum_B \frac{2J_B+1}{2\pi^2} \int_0^\infty \sqrt{k_{B}^2 + m_{B}^{*2}}(\exp[\frac{\varepsilon_B(k) - \mu_B}{T}] + 1)^{-1} k_{B}^2 \mathrm{d}k_{B} \\
&\quad + \sum_l \frac{2J_l+1}{2\pi^2} \int_0^\infty \sqrt{k_{l}^2 + m_l^2} (\exp[\frac{\varepsilon_l(k) - \mu_l}{T}] + 1)^{-1} k_{l}^2 \mathrm{d}k_{l},
\end{aligned}
\end{equation}
\begin{equation}
\begin{aligned}
P &= -\frac{1}{2} m_{\sigma}^{2} \sigma^{2} - \frac{1}{2} m_{\sigma^{*}}^{2} \sigma^{*2}  + \frac{1}{2} m_{\omega}^{2} \omega_{0}^{2}+ \frac{1}{2} m_{\phi}^{2} \phi^{2} + \frac{1}{2} m_{\rho}^{2} \rho_{03}^{2} - \frac{1}{3} a \sigma^{3} - \frac{1}{4} b \sigma^{4} +\frac{1}{4}c_{3}\omega^{4}\\
&\quad + \frac{1}{3} \sum_B \frac{2J_B+1}{2\pi^2} \int_0^\infty \frac{k_{B}^4}{\sqrt{k_{B}^2 + m_{B}^{*2}}} (\exp[\frac{\varepsilon_B(k) - \mu_B}{T}] + 1)^{-1}\mathrm{d}k_{B} \\
&\quad + \frac{1}{3} \sum_l \frac{2J_l+1}{2\pi^2} \int_0^\infty \frac{k_{l}^4}{\sqrt{k_{l}^2 + m_l^2}} (\exp[\frac{\varepsilon_l(k) - \mu_l}{T}] + 1)^{-1}\mathrm{d}k_{l}.
\end{aligned}
\end{equation}
The term $J_{B, l}$ is the spin quantum number. The effective mass of baryons is
\begin{equation}
\begin{aligned}
 m_{B}^{*}=m_{B}-(g_{\sigma B}\sigma + g_{\sigma^* B}\sigma^*).
\end{aligned}
\end{equation}
The condition of charge neutrality requires that
\begin{equation}
\begin{aligned}
q_B \rho_B + q_e \rho_e = 0.
\end{aligned}
\end{equation}
Here $q_B$ is the electric charge of baryon B. The lepton fraction $Y_l = (\rho_e + \rho_{\nu e})/\rho_B$ is fixed at 0.4, a standard value adopted in the literature for PNS matter shortly after birth(e.g.,\cite{Prakash1997,Yu20091}). Moreover, since muons do not appear when neutrinos are trapped, thus $Y_{l\mu} = Y_\mu + Y_{\nu\mu} = 0$. 

\subsection{Structure and Global Properties of PNSs: Mass, Radius, Moment of Inertia, and Surface Gravitational Redshift}
To confront theoretical models with pulsar observations, it is necessary to compute macroscopic stellar properties. With the EOS prescribed, the mass and radius of PNS are derived by the Tolman-Oppenheimer-Volkoff (TOV) equations for hydrostatic equilibrium\cite{Tolman1939,Oppenheimer1939}:
\begin{equation}
\begin{aligned}
\frac{dp}{dr} = -\frac{(p+\varepsilon)(m + 4\pi r^3 p)}{r(r-2m)},\\
m = 4\pi \int_0^r \varepsilon r'^2 dr',\\
M = m(R).
\end{aligned}
\end{equation}

The spacetime metric for a slowly rotating PNS in spherical coordinates$(t, r, \theta, \phi)$ is given by\cite{Hong2018, Zhao2021, Zhao20221, Zhao20222, Zhao2023}
\begin{equation}
\begin{aligned}
ds^2 = & -e^{2\phi(r)} dt^2 + \left[ 1 - \frac{2m(r)}{r} \right]^{-1} dr^2 \\
       & - 2\omega r^2 \sin^2\theta \, dt d\phi + r^2 (d\theta^2 + \sin^2\theta d\phi^2),
\end{aligned}
\end{equation}
At leading order, the moment of inertia of PNS is obtained by
\begin{equation}
\begin{aligned}
I \equiv\frac{J}{\Omega}= \frac{8\pi}{3} \int_0^R (\varepsilon + P) e^{-\phi(r)} \left[ 1 - \frac{2m(r)}{r} \right]^{- \frac{1}{2}} \frac{\varpi}{\Omega} r^4 \mathrm{d}r,
\end{aligned}
\end{equation}
where $\Omega$ denotes the stellar angular velocity, while $\omega$ is the angular velocity measured by a distant observer. 
The quantity $\varpi=\Omega-\omega$ represents the frame-dragging angular velocity, and the angular velocity of a point within the star, as measured by this observer relative to the local inertial frame, is specified by the equation:
\begin{equation}
\begin{aligned}
\frac{1}{r^4} \frac{d}{dr} \left[ r^4 j(r) \frac{d\varpi}{dr} \right] + \frac{4}{r} \frac{dj(r)}{dr} \varpi= 0.
\end{aligned}
\end{equation}
And the metric function $\phi(r)$ determined by
\begin{equation}
\begin{aligned}
\frac{\mathrm{d}\phi(r)}{\mathrm{d}r} = \frac{m(r) + 4\pi r^3 P(r)}{r \left( r + 2m(r) \right)},
\end{aligned}
\end{equation}
where
\begin{equation}
\begin{aligned}
j(r) =
\begin{cases}
\left[1 - \frac{2m(r)}{r} \right]^{\frac{1}{2}} e^{-\phi(r)}, & r \leqslant R, \\[6pt]
1, & r > R.
\end{cases}
\end{aligned}
\end{equation}
The gravitational redshift at the PNS surface can be evaluated using\cite{Douchin2001, Benjamin2006}
\begin{equation}
\begin{aligned}
z = \frac{1}{\sqrt{1 - 2(M/R)}} - 1.
\end{aligned}
\end{equation}
It is evident that the theoretical  moment of inertia and surface gravitational redshift of a PNS can be obtained by solving the TOV equations.
\begin{table}[htbp]
\centering
\small
\setlength{\tabcolsep}{4pt}
\renewcommand{\arraystretch}{1.15}
\caption{The GM1 parameter set \cite{Miyatsu2013, Xu2023}. }
\label{tab:parameter}
\begin{tabular}{@{}lccccccccc@{}}
\hline
\multicolumn{9}{c}{\textbf{Saturation properties of nuclear matter}} \\
\hline
Quantity & $\rho_B^0$ & $K_0$ & $\varepsilon_0$ & $a_4$ & $L$ & $m_N^*/m_N$ \\
Unit     & (fm$^{-3}$) & (MeV) & (MeV) & (MeV) & (MeV) & -- \\
Value    & 0.153 & 300 & $-16.3$ & 32.5 & 93.9 & 0.70 \\
\hline
\multicolumn{10}{c}{\textbf{Coupling constants}} \\
\hline
Coupling & $g_{\sigma\Lambda}$ &$g_{\sigma\Sigma}$ & $g_{\sigma\Xi}$ & $g_{\sigma^*\Lambda}$ & $g_{\sigma^*\Xi}$ & $g_{\omega N}$ & $g_{\sigma^* N}$ & $g_{\phi N}$ \\
Value (SU6) & 5.84 & 3.87 & 3.06 & 3.73 & 9.67 & 10.61 & 0.00 & -- \\
Value (SU3) & 7.25 & 5.28 & 5.87 & 2.60 & 6.82 & 10.26 & 0.00 & $-3.50$ \\
\hline
Coupling  & $g_{\sigma N}$ & $g_{\rho N}$ & $g_{\rho\Lambda}$ & $a$ (fm$^{-1}$) & $b$ \\
Value (SU6) & 9.57 & 4.10 & 0.00 & 12.28 & $-8.98$ \\
Value (SU3) & 9.57 & 4.10 & 0.00 & 12.28 & $-8.98$ \\
\hline
\multicolumn{10}{c}{\textbf{Baryon and meson masses}} \\
\hline
Particle   & $m_\sigma$ & $m_{\sigma^*}$ & $m_\omega$ & $m_\rho$ & $m_\phi$ & $m_N$ & $m_\Lambda$ & $m_\Sigma$ & $m_\Xi$ \\
Mass (MeV) & 550 & 975 & 783 & 770 & 1020 & 939 & 1116 & 1193 & 1318 \\
\hline
\end{tabular}
\end{table}

\subsection{The RMFT coupling constants}
In this work, the GM1 parameter set, widely employed to describe the properties of neutron stars, is adopted for the nucleon coupling constants in PNS matter. They are determined by the saturation properties of nuclear matter, encompassing the saturation density ($\rho_B^0$), incompressibility coefficient ($K_0$), binding energy per nucleon ($\varepsilon_0$), symmetry energy ($a_4$), slope parameter ($L$), and the ratio of the effective nucleon mass to the nucleon mass ($m_{N}^{*}/m_N$)\cite{Glendenning2000,Bednarek2006}. Since the critical density for the onset of hyperons is significantly higher than nuclear saturation density, the nucleon–meson coupling constants cannot be extrapolated to the high-density regime. Instead, the hyperon–meson coupling constants in the work are determined using the widely adopted SU(3) flavor and SU(6) spin–flavor symmetries\cite{Miyatsu2013,Xu2023,Xu2026}. 

The couplings of the $\omega$, $\rho$ and $\phi$ mesons to baryons obey the following relations in SU(3) flavor and SU(6) spin–flavor symmetries: $g_{\sigma^*\Sigma}=g_{\sigma^*\Lambda}$,  $g_{\omega \Sigma}=g_{\omega \Lambda}$, $g_{\rho \Sigma}=2g_{\rho\Xi} = 2 g_{\rho N}$ and $g_{\phi \Sigma}=g_{\phi \Lambda}$.  Under SU(3) flavor symmetry, they can be obtained from 
\begin{equation}
\begin{aligned}
g_{\omega\Lambda} = \frac{g_{\omega N}}{1 + \sqrt{3}z \tan\theta_v}, g_{\omega\Xi} = \frac{(1 - \sqrt{3}z \tan\theta_v)g_{\omega N}}{1 + \sqrt{3}z \tan\theta_v}, \\
g_{\phi\Lambda} = \frac{-\tan\theta_v  g_{\omega N}}{1 + \sqrt{3}z \tan\theta_v}, g_{\phi\Xi} = -\frac{(\sqrt{3}z + \tan\theta_v) g_{\omega N}}{1 + \sqrt{3}z \tan\theta_v}. 
\end{aligned}
\end{equation}
Here, $\theta_v=37.5^\circ$, $z=0.1949$.

UnderSU(6) spin–flavor symmetry, they can be obtained from 
\begin{equation}
\begin{aligned}
g_{\omega\Lambda} = \frac{2}{3}g_{\omega N}, g_{\phi\Lambda} =\frac{1}{2}g_{\phi\Xi}=\frac{\sqrt{2}}{3}g_{\omega N}
\end{aligned}
\end{equation}
Here, $\theta_v \approx 35.26^\circ$, $z=\frac{1}{\sqrt{6}}$. The other specific values of GM1 parameter set are listed in Table 1. 

\begin{figure*}
    \centering
       \includegraphics[width=0.75\linewidth]{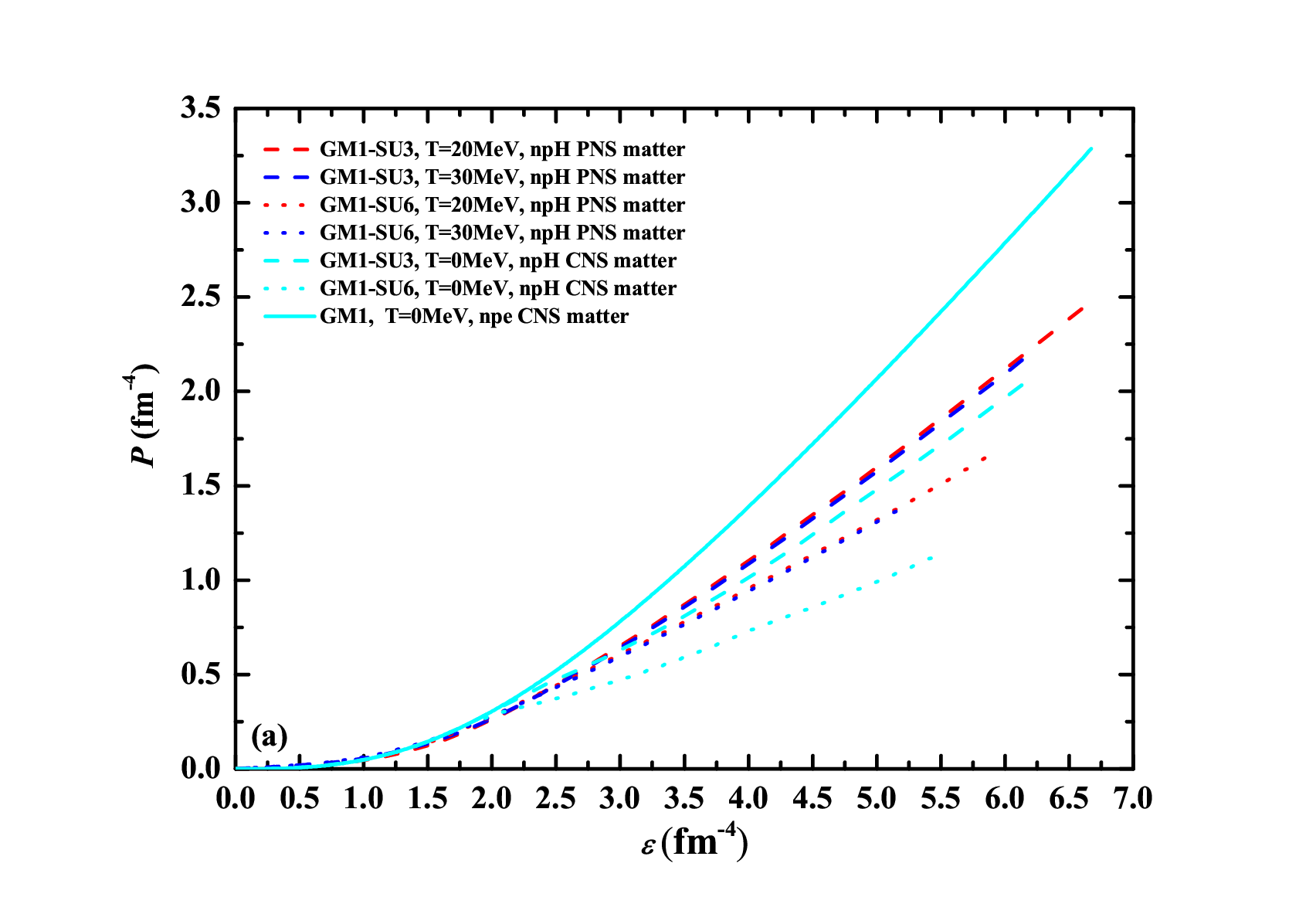}
       \vspace{6pt}

       \includegraphics[width=0.75\linewidth]{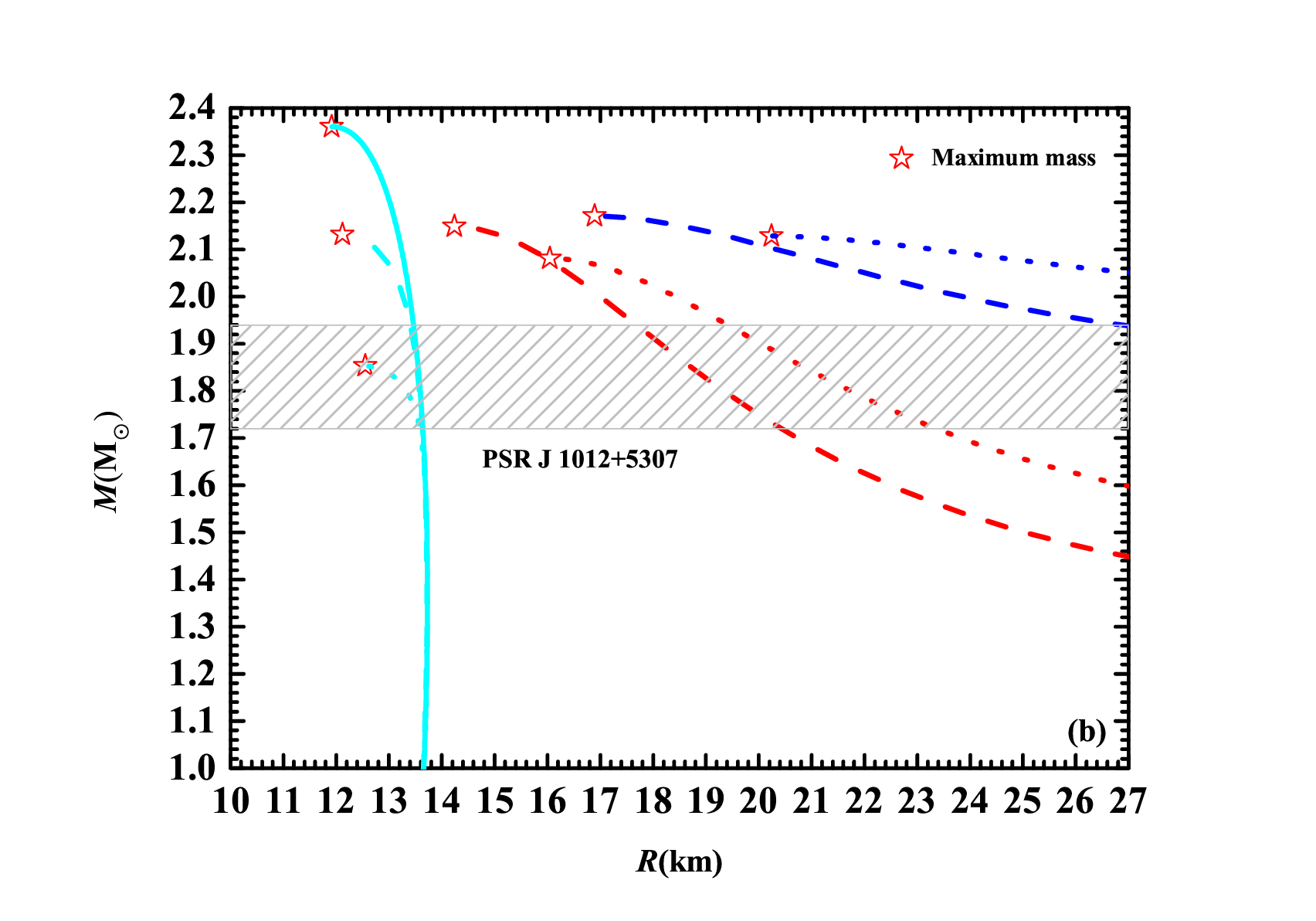}
       \vspace{6pt}

       \includegraphics[width=0.75\linewidth]{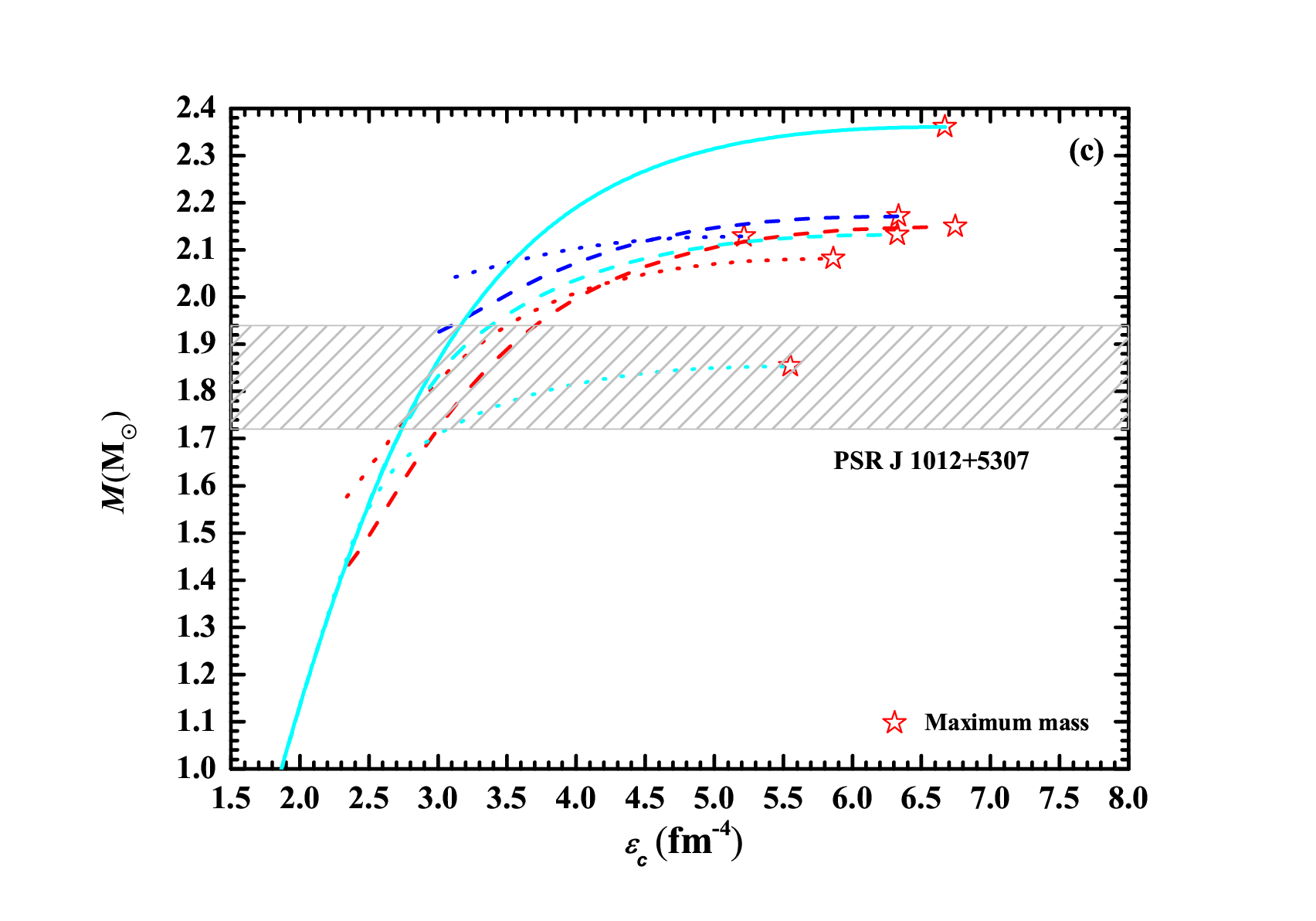}
   
    \caption{
        \textbf{(a)} EOSs of PNSs and CNSs. 
        \textbf{(b)} Mass-radius relations. 
        \textbf{(c)} Mass-central energy density relations. 
        Cyan lines: CNS results at $T=0$~MeV. Solid corresponds to npe matter, dashed and dotted lines correspond to npeH matter with SU(3) flavor and SU(6) spin-flavor symmetries, respectively. 
        Red and blue lines: PNS results at $T=20$~MeV (red) and $T=30$~MeV (blue) under SU(3) (dashed) and SU(6) (dotted) symmetries, respectively. 
        Red stars indicate the maximum masses for each configuration. This notation is used consistently in subsequent figures. }
    \label{fig:eos_mr_mrho_s_rho}
\end{figure*}

\begin{figure*}
    \centering
    \includegraphics[width=0.75\linewidth]{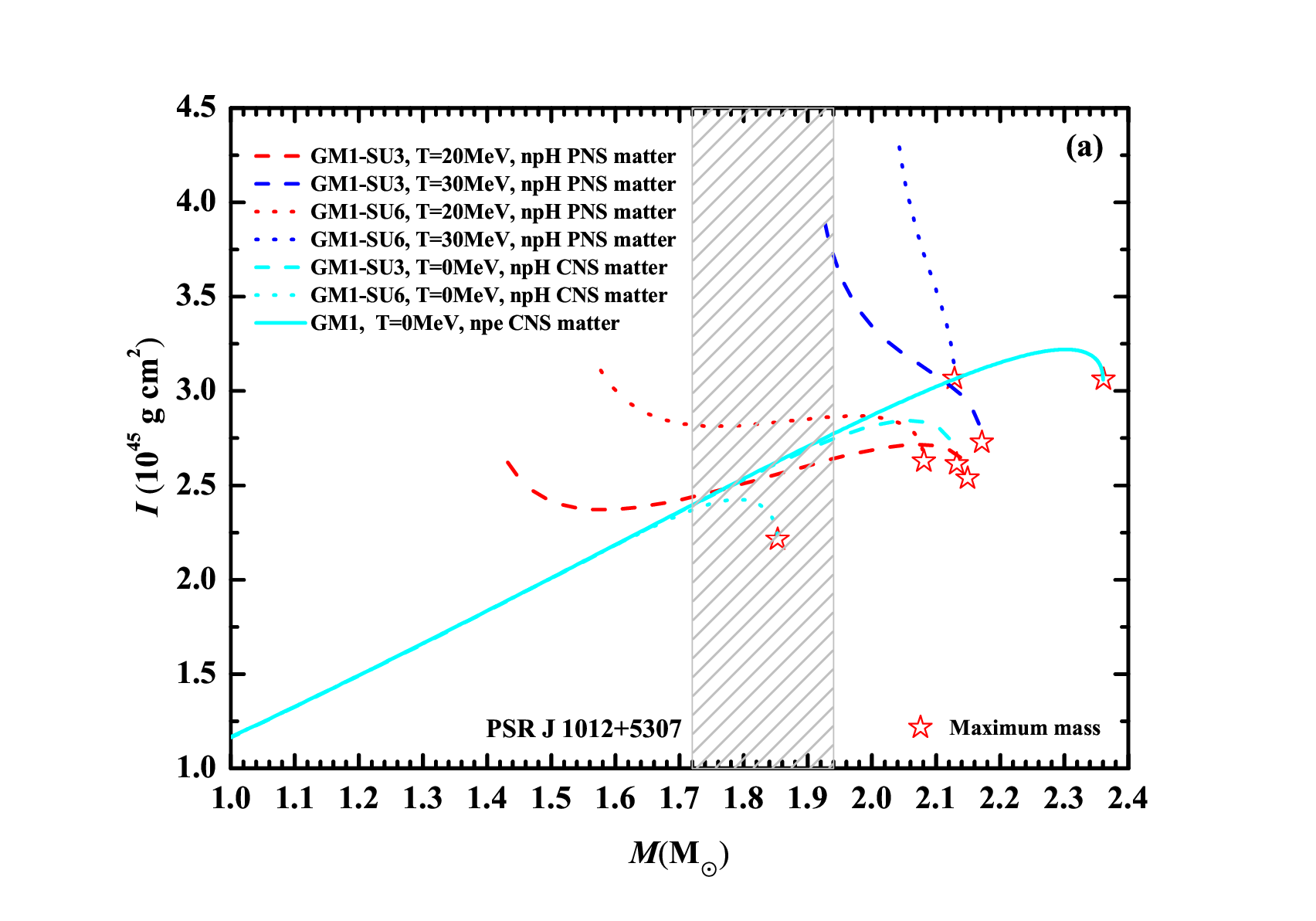}
    \vspace{6pt}
    
    \includegraphics[width=0.75\linewidth]{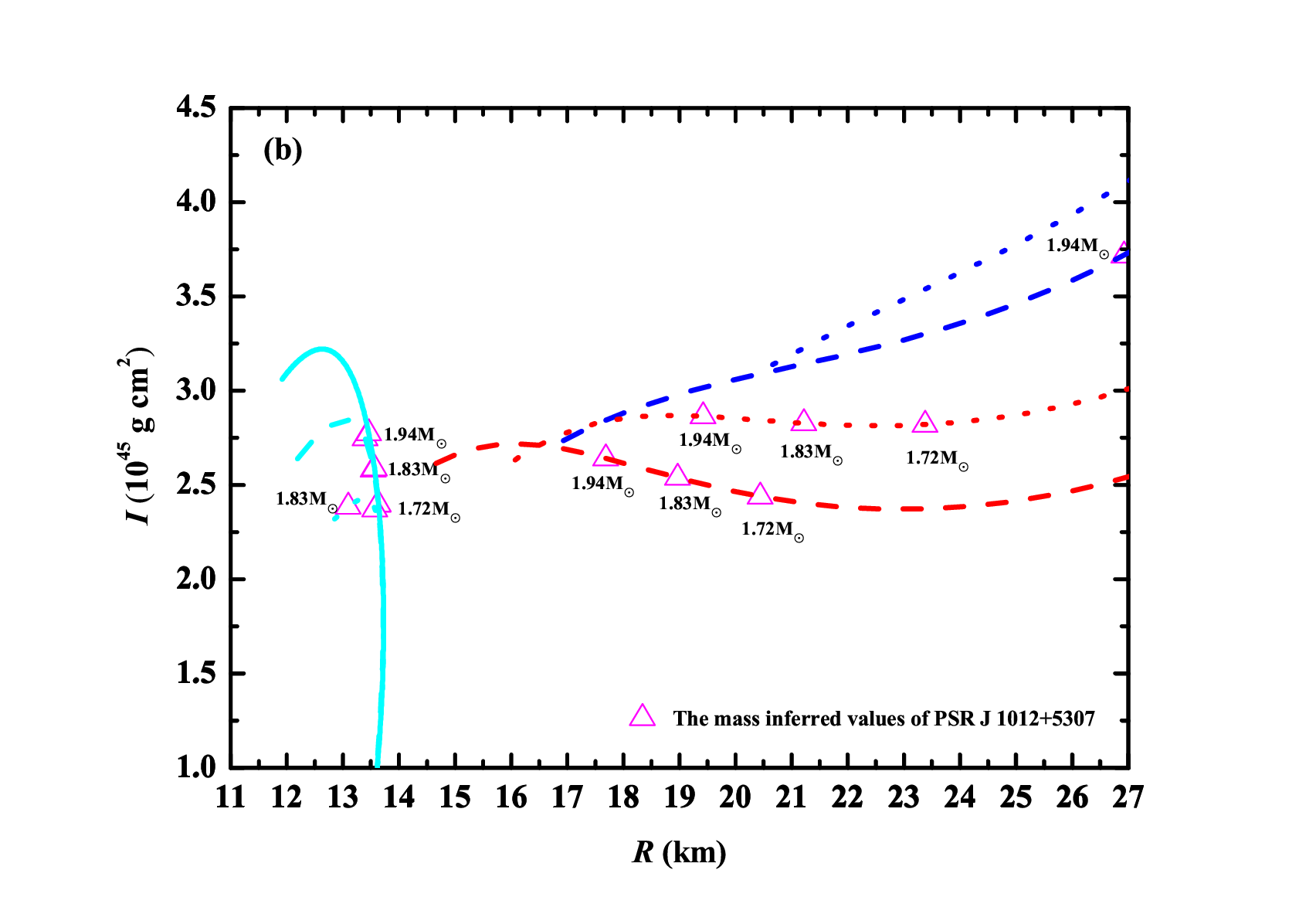}
    \vspace{6pt}
    
    \includegraphics[width=0.75\linewidth]{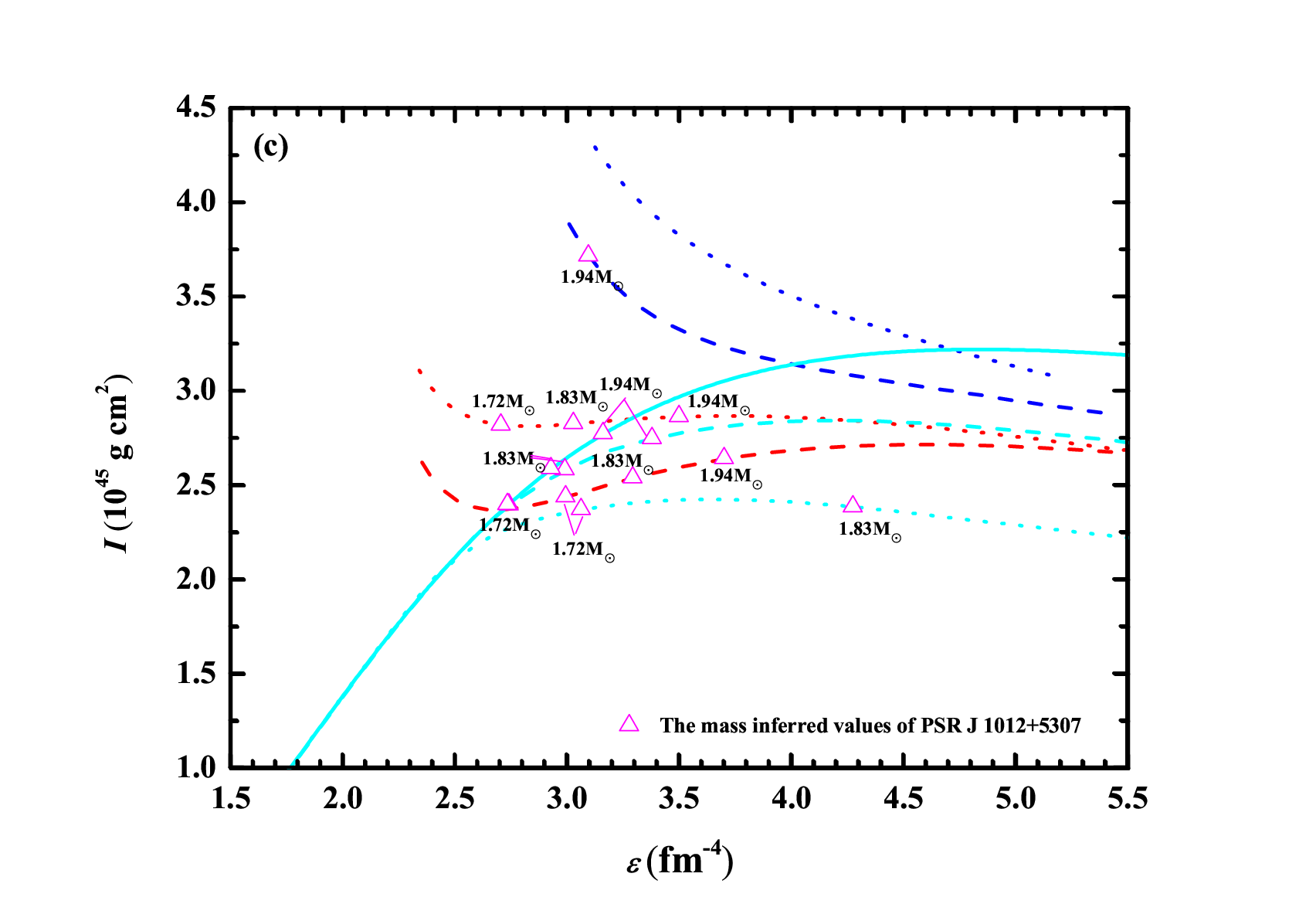}
    
    \caption{
        Moment of inertia of PNSs and CNSs. 
        \textbf{(a)} Moment of inertia as a function of mass. 
        \textbf{(b)} Moment of inertia as a function of radius. 
        \textbf{(c)} Moment of inertia as a function of central energy density. 
        Pink triangles indicate mass measurements of PSR J1012+5307. 
        This notation is used consistently in subsequent figures.
    }
    \label{fig:I_M_R_Ec}
\end{figure*}
\section{Analysis and Discussion}
In this work, we analyze the impact of temperature on the macroscopic properties of PNSs, focusing on the EOSs, mass–radius profiles, mass–energy density relations, and the correlations of moment of inertia and gravitational redshift with mass, radius, and energy density. The analysis is performed at temperatures of $T = 20 $ and 30 MeV, employing the RMFT model with the GM1 parameter set under SU(3) flavor and SU(6) spin-flavor symmetries, respectively. Specifically, we take the intermediate-mass or massive PNS PSR J1012+5307 (measured mass of 1.83$\pm$0.11$ M_{\odot}$)  as a representative case to investigate these macroscopic properties within its specific mass range. 

Fig.~\ref{fig:eos_mr_mrho_s_rho} illustrates the EOSs of PNSs, along with the mass-radius and mass-energy density relationships.​ For comparison, the numerical results for CNSs are shown by the solid, dashed, and dotted cyan lines, representing p, n, e, $\nu_{e}$ (npe) matter, npH matter under SU(3) flavor symmetry, and npH matter under SU(6) spin-flavor symmetry, respectively. Finite-temperature EOSs for hyperonic PNS matter at $T = 20$ MeV are shown by the red dashed (SU(3)) and dotted (SU(6)) lines, while the corresponding results at $T = 30$~MeV are shown by blue dashed (SU(3)) and dotted (SU(6)) lines. Red stars mark the maximum-mass values. This notation is adopted consistently throughout the work.  

As shown in Fig.~\ref{fig:eos_mr_mrho_s_rho}(a), the EOSs exhibit a non-monotonic dependence on temperature. Under both SU(3) flavor and SU(6) spin-flavor symmetries, the EOS at $T=30$ MeV is stiffer than that at $T=20$ MeV at low energy densities, but gradually softens at higher energy densities. For npH matter under SU(6) spin-flavor symmetry, the EOS at $T = 0$ MeV is slightly softer than that at $T = 20/30$ MeV at low energy densities; however, as the energy density increases, it first stiffens progressively before softening considerably, leading to an increasingly significant deviation from the finite-temperature PNS curves at intermediate and high energy densities. For npH matter under SU(3) flavor symmetry, the $T = 0$ MeV EOS exhibits a similar behavior to that observed under SU(6) spin-flavor symmetry at low densities when compared to the $T = 20/30$ MeV cases. Nevertheless, with increasing energy density, this evolution is notably more moderate compared to the  temperature dependence observed under SU(6) spin-flavor symmetry. This indicates that the distinct variations in meson-baryon coupling strengths between the SU(3) flavor and SU(6) spin-flavor symmetries$-$particularly the repulsive contribution of the $\phi$ meson to the nucleon–nucleon interaction (see Table~\ref{tab:parameter})$-$result in a significantly smaller temperature-induced deviation in the EOS for SU(3) compared to the SU(6) case. Furthermore, at a fixed temperature ($T = 0$, 20 or 30 MeV), the EOS of npH matter under SU(3) flavor symmetry remains significantly stiffer than that under SU(6) spin-flavor symmetry in the intermediate- to high-energy density regime. Notably, the CNS composed solely of npe matter retains the stiffest EOS. Overall, the temperature dependence of the EOSs for hyperonic PNSs remains sufficient to drive​ substantial changes in macroscopic properties, such as the mass-radius and mass-energy density relations illustrated in Fig.~\ref{fig:eos_mr_mrho_s_rho}(b) and Fig.~\ref{fig:eos_mr_mrho_s_rho}(c).  Specifically, as shown in Fig.~\ref{fig:eos_mr_mrho_s_rho}(b) and Fig.~\ref{fig:eos_mr_mrho_s_rho}(c), the radius of a PNS at a fixed mass decreases markedly with decreasing temperature, whereas the central energy density rises, regardless of the symmetry considered. For fixed radius or energy density, the PNS mass exhibits a significant decreasing trend with decreasing temperature under both SU(3)  flavor and SU(6) spin-flavor symmetries. To clearly illustrate the macroscopic properties of maximum-mass stars and PSR J1012+5307, Tables~\ref{tab:max} and \ref{tab:psr_j1012_5307} list the radius, central energy density, moment of inertia, and gravitational redshift for these configurations under both symmetries.

\begin{figure*}
    \centering
    \includegraphics[width=0.75\linewidth]{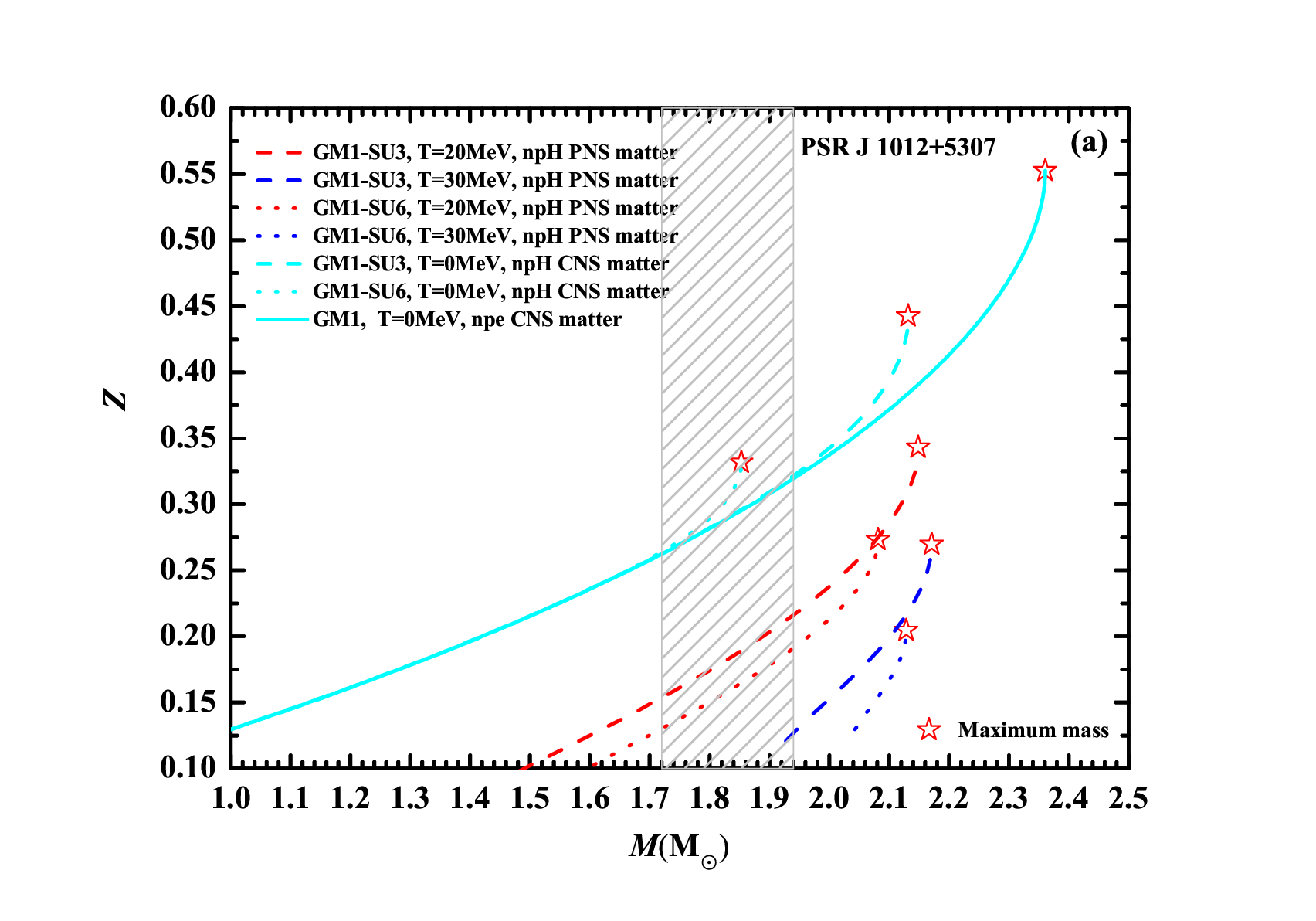}
   \vspace{6pt}

    \includegraphics[width=0.75\linewidth]{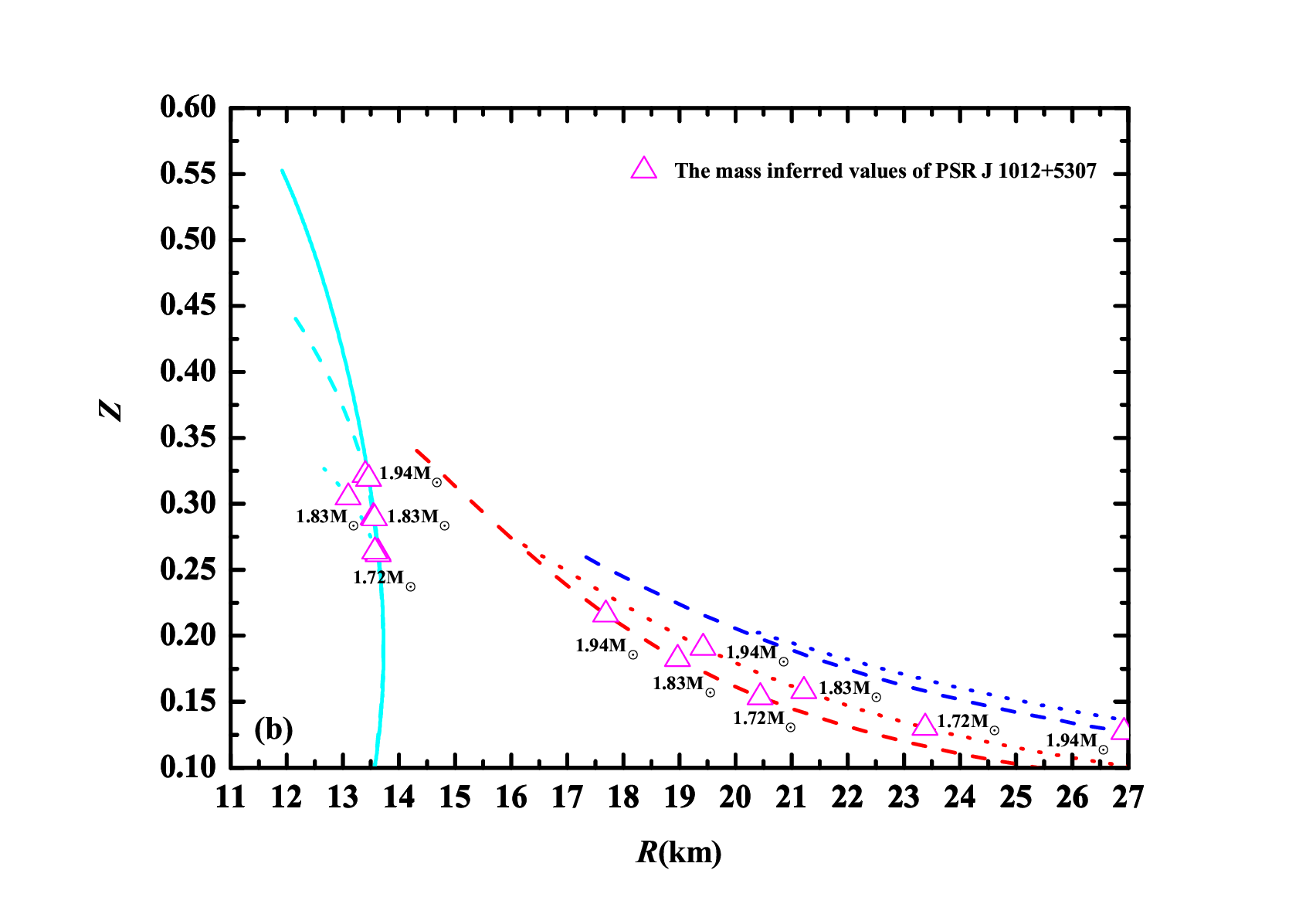}
    \vspace{6pt}

   \includegraphics[width=0.75\linewidth]{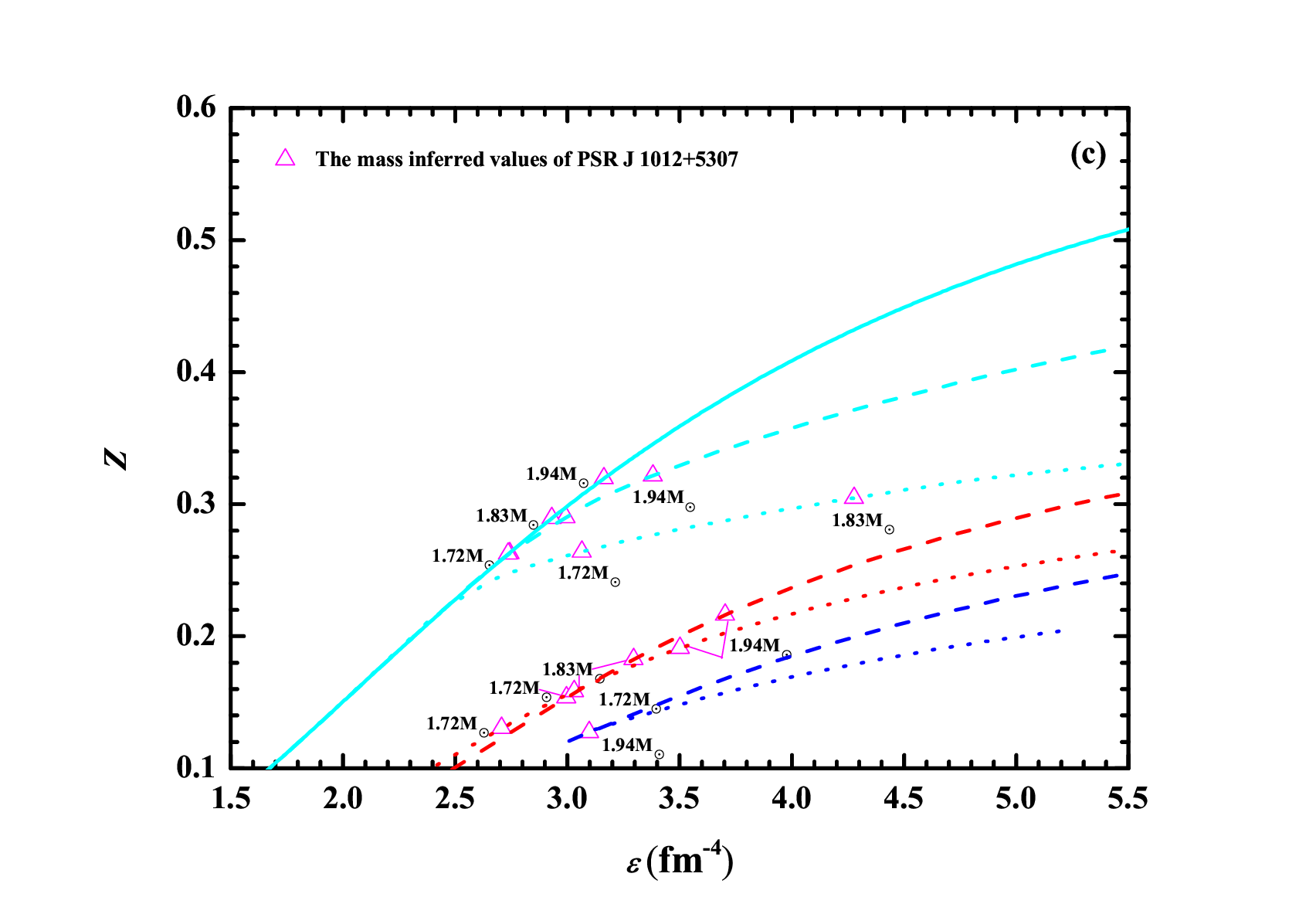}
   
    \caption{
        Gravitational redshift of PNSs and CNSs. 
        \textbf{(a)} Gravitational Redshift as a function of mass. 
        \textbf{(b)} Gravitational Redshift as a function of radius. 
        \textbf{(c)} Gravitational Redshift as a function of central energy density.
    }
    \label{fig:Z_M_R_Ec}
\end{figure*}

The macroscopic properties of PNSs and CNSs, particularly the moment of inertia ($I$) and gravitational redshift ($z$), manifest distinct structural responses to temperature variations, as illustrated in Figs.~\ref{fig:I_M_R_Ec} and \ref{fig:Z_M_R_Ec}. For CNSs ($T = 0$~MeV), $I$ generally increases with mass but declines in the high-mass regime, consistent with standard stellar structure theory. Under both symmetry schemes, $I$ of the PNS initially decreases with increasing mass at $T = 20$ MeV.  At higher masses, however, its evolution closely follows that of a CNS. In contrast, at $T = 30$ MeV, $I$ exhibits a monotonic decline across the entire mass range. Conversely, $z$ increases monotonically with mass for both CNSs and PNSs. Moreover, $z$ of the CNS consistently exceeds that of the PNS at a given mass. Furthermore, $I$ and $z$ of the PNS decrease with decreasing temperature, regardless of whether the radius ($R$) or the central energy density ($\varepsilon$) is held fixed. For npH matter under SU(3) flavor symmetry, $I$ of the PNS exceeds that of the CNS at $T=30$ MeV across the entire mass range, whereas at $T=20$ MeV, $I$ of the PNS is larger than that of the CNS at low and intermediate masses but falls below it at higher masses. This trend is particularly evident for PSR J1012+5307. Based on the numerical results presented in Figs.~\ref{fig:I_M_R_Ec}(b) and \ref{fig:I_M_R_Ec}(c), \ref{fig:Z_M_R_Ec}(b) and \ref{fig:Z_M_R_Ec}(c), and Table~ \ref{tab:psr_j1012_5307},​ ​we now examine the temperature-driven transition from a PNS to a CNS for hyperonic PSR J1012+5307 under SU(3) flavor symmetry. Taking a 1.94 $ M_{\odot}$ configuration as an example, decreasing the temperature from $T = 30$ MeV to 20 MeV and finally to 0 MeV results in a dramatic $R$ contraction from 26.917 km to 17.686 km, ultimately stabilizing at 13.400 km. This contraction is accompanied by a rise in $\varepsilon$ from 3.097 fm$^{-4}$ to 3.704 fm$^{-4}$ at $T = 20$ MeV, before settling at 3.382 fm$^{-4}$ in the cold phase. Concurrently, $I$ drops from 3.718$\times10^{45}$ g cm$^2$  to 2.641$\times10^{45}$ g cm$^2$,  eventually recovering slightly to 2.747$\times10^{45}$ g cm$^2$. In tandem, $z$ deepens significantly from 0.127 to 0.216, reaching 0.322 in the cold phase. In addition, for npH matter under SU(6) spin-flavor symmetry,  $I$ of the PNS for both T=20 and 30 MeV remains larger than that of the CNS.
\begin{table}[htbp]
  \centering
\small
  \setlength{\tabcolsep}{3pt} 
  \caption{Radius, central energy density, moment of inertia, and gravitational redshift for maximum-mass PNSs and CNSs under SU(3) flavor and SU(6) spin-flavor symmetries.}
  \begin{tabular}{|l|c|c|c|c|c|c|}
    \hline
    Parameter (Unit) & $T$ (MeV) & $M_{\text{max}}$ ($M_{\odot}$) & $R$ (km) & $\varepsilon_{0}$ (fm$^{-4}$) & $I$ ($10^{45}$ g cm$^2$) & $z$ \\
    \hline
    \multirow{4}{*}{npeH (SU(3))} & 0 & 2.132 & 12.119 & 6.326 & 2.613 & 0.443 \\
    \cline{2-7}
    & 20 & 2.149 & 14.241 & 6.746 & 2.537 &0.343 \\
    \cline{2-7}
    & 30 & 2.171 & 16.890 & 6.334 & 2.727 & 0.270 \\
    \hline
    \multirow{4}{*}{npeH (SU(6))} & 0 & 1.853 & 12.549 & 5.552 & 2.214 & 0.332 \\
    \cline{2-7}
    & 20 & 2.081 & 16.045 & 5.862 & 2.627 & 0.273 \\
    \cline{2-7}
    & 30 & 2.129 & 20.238 & 5.216 & 3.065& 0.204 \\
    \hline
    npe & 0 & 2.361 & 11.915 & 6.671 & 3.061 & 0.552 \\
    \hline
  \end{tabular}
  \label{tab:max}
\end{table}

\begin{table}[htbp]
  \centering
 \small
  \setlength{\tabcolsep}{1pt} 
  \caption{Radius, central energy density, moment of inertia, and gravitational redshift for CNSs and PNSs within the mass range of PSR J1012+5307, calculated at various temperatures under SU(3) flavor and SU(6) spin-flavor symmetries.}
  \begin{tabular}{|l|c|c|c|c|c|c|}
    \hline
  \multirow{2}{*}{PSR J-1012+5307}& \multicolumn{6}{c|}{SU(3) (SU(6))} \\
    \cline{2-7}
  \multirow{12}{*}{npeH}& $T$ (MeV) & $M$ ($M_{\odot}$) & $R$ (km) & $\varepsilon_{0}$ (fm$^{-4}$) & $I$ ($10^{45}$ g cm$^2$) & $z$ \\
    \cline{1-7}
    & 0 & 1.72 & 13.625(13.576) & 2.744(3.065) & 2.395(2.371) & 0.263(0.264) \\
    \cline{2-7}
    & 0 & 1.83 & 13.543(13.099) & 2.992(4.277) & 2.581(2.386) & 0.290(0.305) \\
    \cline{2-7}
    & 0 & 1.94 & 13.400(-) & 3.382(-) & 2.747(-) & 0.322(-) \\
    \cline{2-7}
    & 20 & 1.72 & 20.435(23.375) & 2.997(2.708) &2.440(2.820)  & 0.154(0.130) \\
    \cline{2-7}
    & 20 & 1.83 & 18.966(21.214) & 3.296(3.030) & 2.538(2.827) & 0.183(0.158) \\
    \cline{2-7}
    & 20 & 1.94 & 17.686(19.417) & 3.704(3.502) & 2.641(2.864) & 0.216(0.191) \\
    \cline{2-7}
    & 30 & 1.72 & -(-) & -(-) & -(-) & -(-) \\
    \cline{2-7}
    & 30 & 1.83 & -(-) & -(-) & -(-) & -(-) \\
    \cline{2-7}
    & 30 & 1.94 & 26.917(-) & 3.097(-) & 3.718(-) & 0.127(-) \\
    \cline{1-7}
    \multirow{3}{*}{npe} & 0 & 1.72 & 13.636 & 2.736 & 2.397 & 0.262 \\
    \cline{2-7}
    & 0 & 1.83 & 13.564 & 2.930 & 2.588 & 0.289 \\
    \cline{2-7}
    & 0 & 1.94 & 13.461 & 3.163 & 2.774 & 0.319 \\
    \hline
  \end{tabular}
  \label{tab:psr_j1012_5307}
\end{table}
If this object is a hot PNS containing hyperons (npH matter), we can examine the variations​ in $I$ and $z$ arising from mass uncertainties. Taking the T=20 MeV case as an example, under SU(3) flavor symmetry, increasing the mass from 1.72 $M_{\odot}$ to 1.94 $M_{\odot}$ corresponds to $R$ contraction from 20.435 km to 17.686 km, accompanied by a rise in $\varepsilon$ from 2.997 fm$^{-4}$ to 3.704 fm$^{-4}$. Concurrently, $I$ increases from 2.440$\times 10^{45}$ g cm$^2$ to 2.641$\times10^{45}$ g cm$^2$, while $z$ increases from 0.154 to 0.216. Analogous behavior is observed under SU(6) spin--flavor symmetry. Further refining the mass measurement of this star would enable more detailed studies to help determine whether hyperons appear in a PNS and identify their specific types. Interestingly, as listed in Table~ \ref{tab:psr_j1012_5307}, for PSR J1012+5307 at $T = 0$ MeV, the macroscopic properties of npH matter under SU(3) flavor symmetry differ only marginally from​ those of npe matter. For instance, at a fixed mass of 1.94 $ M_{\odot}$ under SU(3) flavor symmetry, the npH star has $R=13.400$ km compared to​ 13.461 km for the npe star; $\varepsilon=$ 3.382 fm$^{-4}$ vs. 3.163 fm$^{-4}$; $I=$ 2.747$\times 10^{45}$ g cm$^2$ vs. 2.774$\times 10^{45}$ g cm$^2$; and $z=$ 0.322 vs.​ 0.319.  A similar pattern holds  at 1.72 and 1.83 $M_{\odot}$, where the discrepancies remain equally small. These marginal differences imply that observationally confirming the presence of hyperons in cold pulsars like PSR J1012+5307 remains challenging with current data. However, this challenge may be overcome by tracking the evolutionary history. Indeed, if future observations can track the thermal evolution of a pulsar from its birth through the PNS-to-CNS transition, the variations in radius, moment of inertia, and gravitational redshift predicted by our models may provide valuable insights into exotic matter. While this work focuses on hyperonic EOSs within the RMFT framework using the GM1 parameter set, it provides a foundation for understanding how thermal effects reveal strange particles in neutron stars. Future studies incorporating additional exotic degrees of freedom, such as quark matter, will further refine these constraints.

\section{summarizes}
In summary, this study highlights the crucial role of temperature in determining the macroscopic properties of hyperonic PNSs. We find that the EOS exhibits a non-monotonic temperature dependence, leading to a substantial structural transformation during the transition from a PNS to CNS. Taking the case of  the PNS of PSR J1012+5307 with a mass of 1.94 $ M_{\odot}$ under SU(3) flavor symmetry, we demonstrate that the star undergoes a drastic contraction as the temperature decreases from $T = 30$ MeV to 0 MeV. The radius contracts by approximately 50$\%$ (from 26.917 km to 13.400 km), accompanied by a reduction in the moment of inertia by nearly 26$\%$ (from 3.718 $\times 10^{45}$ g cm$^2$ to 2.747$\times 10^{45}$ g cm$^2$ ), and an increase in the gravitational redshift by approximately 154$\%$ (from 0.127 to 0.322). To quantify the impact of mass uncertainty for the PNS of PSR J1012+5307, we analyze the variations in the moment of inertia and gravitational redshift at $T=20$ MeV. Taking SU(3) symmetry as an example, increasing the mass across the range 1.72$ M_{\odot}$$-$1.94 $ M_{\odot}$ induces a radius contraction of 2.749 km (from 20.435 km to 17.686 km), a $\sim 8\%$ increase in the moment of inertia (from 2.440 $\times 10^{45}$ g cm$^2$ to 2.641$\times 10^{45}$ g cm$^2$ ), and a $\sim 40\%$ rise in the gravitational redshift (from 0.154 to 0.216). Meanwhile, our results reveal that the macroscopic properties of npH and npe matter exhibit only minimal variations​ for PSR J1012+5307 at fixed masses under SU(3) flavor symmetry, rendering it challenging to observationally confirm the presence of hyperons in CNS. Refining the mass measurement of PSR J1012+5307 via future observations, coupled with long-term tracking of a pulsar's evolution from birth, would enable a deeper investigation into the emergence of hyperons and the specific types of exotic matter within PNSs.

\end{document}